\documentclass[preprints,article,accept,moreauthors,dvi2pdf]{mdpi}

\firstpage{1} 
\pubvolume{xx}
\issuenum{1}
\articlenumber{5}
\pubyear{2020}
\copyrightyear{2020}
\history{Received: date; Accepted: date; Published: date}

\pdfoutput=1

\usepackage{graphicx}
\usepackage{dcolumn}
\usepackage{bm}
\usepackage{hyperref}
\usepackage{braket}
\usepackage[title]{appendix}
\newcommand{\bigCI}{\mathrel{\text{\scalebox{1.07}{$\perp\mkern-10mu\perp$}}}}
\Title{Towards a Framework for Observational Causality From Time Series: When Shannon Meets Turing}

\Author{David Sigtermans $^{1 *}$} 

\AuthorNames{David Sigtermans}

\address{%
$^{1}$ \quad ASML}

\corres{Correspondence: david.sigtermans@asml.com}

\abstract{We propose a novel tensor-based formalism for inferring causal structures from time series. An information theoretical analysis of \emph{transfer entropy (TE)}, shows that \emph{TE} results from transmission of information over a set of communication channels. Tensors are the mathematical equivalents of these multi-channel causal channels. A multi-channel causal channel is a generalization of a \emph{discrete memoryless channel (DMC)}. Investigation of a system comprising three variables shows that in our formalism, bivariate analysis suffices to differentiate between direct and indirect relations. For this to be true, we have to combine the output of multi-channel causal channels with the output of single-channel causal channels. We can understand this result when we consider the role of noise. Subsequent transmission of information over noisy channels can never result in less noisy transmission overall. This implies that a \emph{Data Processing Inequality (DPI)} exists for transfer entropy.}


\begin{document}

\section{Introduction}
\subsection{Motivation and Significance of the Work}
Exact knowledge about the causal relationships that determine the behavior of complex systems is a holy grail in the (applied) sciences and engineering. This knowledge enables us to determine potential causes of certain effects, and it allows us to predict the effect of changes in causes (for example via simulation). In other words, it allows for causal inference and causal discovery \cite{Survey}. In a causal relation, the cause precedes the effect (temporal precedence), and the cause physically influences the effect \cite{Eichler}. A causal description is essentially different from a description via statistical associations as illustrated by the adage ``\textit{correlation does not imply causation}’’. Examples of wrong, expensive, or even worse, harmful conclusions and policies based on statistical associations are part of common lore.

For a causal description, intervention is required \cite{Pearl}. These interventions enable us to differentiate between direct and indirect, or spurious, associations \footnote{An indirect association is an association via one or more mediators.}. Because interventions are not always possible, we have to make do with observational data. A plethora of methods to infer causal structures from observational data have been developed, see for example \cite{GrangerORI, TDMI, DBN, Spirtes2000, Schreiber, Lizier2010, ICA-LiNGAM, Runge, DTE, CausationEntropy}. What most these methods have in common is that they express relations via the causal effect, i.e., point-wise estimators that characterize ``strength’’ of the association between a cause and an effect. 

We propose a novel approach inspired by Turing machines \cite{Turing}. If a human ``computer’’ can decide, given the data, if a relation is causal, a Turing machine exists that reaches this decision in a mechanical way \cite{Church_Turing}. This is not a tautology, a Turing machine encodes the underlying principles leading to the decision that a relation is causal. It closely relates our approach to Structural Causal Models \cite{SCM}. Instead of using point-wise estimators like transfer entropy \cite{Schreiber} or time-delayed mutual information \cite{TDMI}, we use stochastic tensors, i.e. multilinear maps. 

Under the assumption that there are no hidden causes, unmeasured common causes or confounders, our formalism can differentiate between direct and indirect associations. We show that noise has a fundamental and, from the viewpoint of detecting spurious associations, a functional role. It is as if noise acts like ``soft’’ interventions \cite{Interventions}. A surprising result is that for time-delayed mutual information \emph{bivariate analysis suffices} to differentiating between direct and indirect associations. This result contradicts the long-held belief that this is impossible (see for example, \cite{Grangercausality} and \cite{Triangle}). The formalism furthermore allows for a simple proof of a Data Processing Inequality \cite{ThomasCover} for transfer entropy. This DPI can identify potential indirect relations when using TE, see for example \cite{ARACNE_DPI}.

\subsection{Outline}
The proposed formalism relies heavily on probability theory \cite{LoTP} and aspects and terminology used in causal inference \cite{Pearl}. In Section \ref{Section: Preliminaries} we give a short overview of the most important ones. 

To derive our formalism, we apply concepts from information theory \cite{Shannon} to transfer entropy. Transfer entropy is a measure that \emph{can} capture causal relations, as far as encoded in the probability density functions, see for example \cite{EffectivenessCI, RASHIDI}. In Section \ref{section:Information theory} we therefore introduction the applicable aspects of information theory, e.g., transmission of information, mutual information, communication channels and the tensor representation of communication channels. A tensor is a multilinear map that transforms an input into an output. We introduce transfer entropy in Section \ref{section:Transfer entropy}. We then show that transfer entropy is the average mutual information resulting from transmission of information of a set of communication channels. We call this set of channels \emph{multi-channel causal channels}. Tensors are the mathematical equivalent to the set of channels. Using these tensors, we establish calculation rules in Section \ref{section: calculation rules}. We restrict ourselves to a system comprising three variables. We derive these results using index notation in Section \ref{section: calculation rules}. The result allow for a different notation that helps us to avoid a notational jungle of indices. This notation is borrowed from quantum mechanics and we introduce it in Section \ref{section: a new notation}. This let us incorporate the temporal relations between a source and a destination, reflecting the additivity of interaction delays. Using this notation, we discuss some relevant findings for causal inference in Section \ref{section: structures with Causal Tensors}. 

In Section \ref{section: Experiments} we present two experiments that illustrate that our formalism can indeed detect nonlinear relationships and an underlying structure.

For readability, we have moved the longer proofs or sketches of proofs to the Appendix.

\subsection{Preliminaries} \label{Section: Preliminaries}
Statistical independence is foundational to causal inference \cite{Spirtes2000}. We will summarize the two most related and relevant assumptions: (1) the faithfulness assumption. (2) the Causal Markov Condition. A directed graph is said to be faithful to the underlying probability distributions if the independence relations that follow from the graph are the same independence relations that follow from the underlying probability distributions. For the chain $X \rightarrow Y \rightarrow Z$ the faithfulness assumption implies that $X$ and $Z$ are independent given $Y$. We denote this as $X \bigCI Z \vert Y$.

The Causal Markov Condition states that a process is independent of its non-effects given its direct causes or its parents. This is relevant in the context of time series. A straightforward interpretation of this condition is that if the set of variables blocks all (undirected) paths between two variables, these two variables are independent given the set of variables blocking all paths \cite{Pearl}. We illustrate the Causal Markov Condition with an example that will be used later in this article.

\begin{figure}[H]
\centering
\includegraphics[width=10 cm]{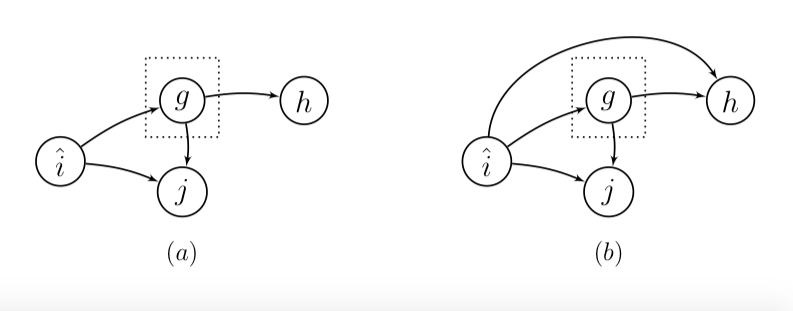}
 \caption{\label{Example1} The graphs used in Example \ref{ex:CausalM}. (\textbf{a}) The dotted box blocks all paths between the non-parents $\hat{i}$ and $j$ of $h$; this implies that $\hat{i}$ and $h$ are independent given $g$. (\textbf{b}) The dotted box does not block the path between $\hat{i}$ and $h$: there is still a path between $\hat{i}$ and $h$. Therefor $\hat{i}$ and $h$ are not independent given $g$}
\end{figure}

\begin{Example}\label{ex:CausalM}
Lets start with the graph depicted in Figure \ref{Example1}(\textbf{a}). According to the Causal Markov Condition, $\{ \hat{i}, j \}$ and $h$ are independent given $g$: $p(\hat{i}, j, h \vert g)\! =\! p(\hat{i},j\vert g)p(h\vert g)$.

We now rewrite this using expressions that follow from the definition for conditional probabilities. The left-hand side is written as $p(\hat{i}, j, h \vert g)\! =\! p(j, h \vert \hat{i},g)p(\hat{i} \vert g)$. The right-hand side can be rewritten using $p(\hat{i},j\vert g)\! =\! p(j\vert \hat{i},g)p(\hat{i}\vert g)$. 

This finally leads the conclusion that $p(j,\vert \hat{i},g,h)\! =\! p(j\vert \hat{i},g)$. This last expression also implies that $p(j\vert \hat{i},g)p(g\vert \hat{i},h) = p(j,g\vert \hat{i},h)$. Both expressions will be used later in this article.\\

Now consider the situation depicted in Figure \ref{Example1}(\textbf{b}). According to the Causal Markov Condition $\{ \hat{i}, j \}$ and $h$ are \emph{not} independent given $g$, i.e., $p(\hat{i}, j, h \vert g)\! \neq \! p(\hat{i},j\vert g)p(h\vert g)$. We can still rewrite the left-hand side and the right-hand side in the same fashion as before. 

This finally leads the conclusion that $p(j,\vert \hat{i},g,h)\! \neq\! p(j\vert \hat{i},g)$. This implies that $p(j\vert \hat{i},g)p(g\vert \hat{i},h) \neq p(j,g\vert \hat{i},h)$.
\end{Example}
 
In the example above we used a simplified notation for the probabilities $p(y)\! :=\! Pr\{Y\! =\! y\}$. As illustrated in the above example we use basic aspects of probability theory like the definition of joint probabilities and the Law of Total Probability \cite{LoTP}. This law links a marginal probability to a joint probability, e.g., $\sum_g p(j,g\vert i,h)\! =\! p(j\vert i,h)$. 

Unless stated otherwise, in this article we will make use of the Einstein summation convention (with a twist). This convention simplifies equations by implying summation over indices that appear both as upper indices and as lower indices. In our definition the summation takes place the first indices and the subsequent identical lower indices $B^{i}_j A^{j}_i := \sum_i B^{i}_j A^{j}_i$. Our definition implies that $B^{i}_j A^{j}_i \neq A^{j}_i B^{i}_j$, the order matters. 

\section{Information Theory} \label{section:Information theory}
Shannon introduced information theory in 1948 \cite{Shannon}. It models association between random variables as resulting from a communication process between a sender---the source---and a receiver or the destination. A message comprises indexed realizations of random variables representing stationary ergodic processes. An input message is first encoded: we describe the message using a finite alphabet. Each random variable has its own finite alphabet. The random variable $X$ is mapped on symbols from the alphabet $\mathcal{X}$, the random variable $Y$ is mapped on symbols from $\mathcal{Y}$, and the random variable $Z$ is mapped on symbols from $\mathcal{Z}$. Where \normalsize $\mathcal{X}\! =\! \{ \chi_1, \chi_2,\cdots, \chi_{\vert \mathcal{X} \vert} \}$\normalsize, $\mathcal{Y}\! =\! \{ \psi_1, \psi_2,\cdots$\normalsize, $\psi_{\vert \mathcal{Y} \vert} \}$\normalsize, and \normalsize$\mathcal{Z}\! =\! \{ \zeta_1, \zeta_2,\cdots, \zeta_{\vert \mathcal{Z} \vert} \}$\normalsize. The number of elements in the alphabet---the cardinality---is denoted as $\vert \mathcal{X} \vert$, $\vert \mathcal{Y} \vert$, and $\vert \mathcal{Z} \vert$ respectively. 

Once encoded the message is transmitted symbol by symbol. The input symbol is transformed into an output symbol. The output alphabet can have a different cardinality than the input alphabet. The transformation from input to output symbol is modeled as a Markov chain. The probability that a specific output symbol is received only depends on the alphabet symbol that was sent. The communication process transforms the input probability mass function (pmf) into the output pmf. The transmitted message is decoded and made available to the receiver. In this article, we assume that no decoding takes place.

\subsection{Mutual Information}
If there is an association between two messages, information is said to be shared between them. The amount of information shared,
 
\begin{equation} \label{eq:MI}
 I(X;Y) = \sum_{x \in \mathcal{X} ,y\in \mathcal{Y}} p(x,y) \log_2 \left[ \frac{p(y|x)}{p(y)} \right],
\end{equation}

is nonnegative and symmetric in $X$ and $Y$. This so-called mutual information (MI) represents the reduction in uncertainty about the random variable $X$ given that we have knowledge about the random variable $Y$ (and vice versa). It is intuitively clear that, given the information content of the source data, in subsequent transmissions, the information can never increase. This is formalized in the Data Processing Inequality or DPI which states that processing of data can never increase the amount of information \cite{ThomasCover}. For the cascade $X\!\rightarrow \! Y \!\rightarrow \! Z$ the DPI implies that, in terms of MI,
 
\begin{equation*} \label{eq:DPI}
 I(X;Z)\! \leq\! \min [I(X;Y),I(Y;Z)].
\end{equation*}

The maximum rate with which information can be transmitted between the sender and receiver is the channel capacity $C_{XY}\! =\! \max_{p(x)} \left[ I(X;Y) \right]$. This is achieved for a so-called channel achieving input distribution.  

\subsection{The Communication Channel}
In information theory, the directed graph representing a Markov chain is represented as a \textit{communication channel}, or channel in short. The channel has an input side---the left-hand side---and an output side---the right-hand side. On the left-hand side we place all the vertices of the Markov chain with outgoing edges and on the right-hand side we place all the vertices of the Markov chain with incoming edges. The input vertices are connected to the output vertices via undirected edges. In a channel, every input alphabet symbol has its own input vertex. Likewise, every output alphabet symbol has its own output vertex. The simplest type of channel is the noisy discrete memoryless communication channel. In a memoryless channel the output---$y_t$---only depends on the input---$x_t$---and not on the past inputs or outputs: $p(y_t \vert x_t, x_{t-1}, y_{t-1})\! =\! p(y_t \vert x_t)$. A memoryless channel embodies the Markov property. In a noisy channel the output depends on the input and another random variable representing noise. The effect of transmitting data using a DMC is described via the Law of Total Probability because

\begin{equation} \label{eq:LawOfTotalProbability}
  Pr\{Y=\psi_{j}\}= \sum_{i} Pr\{X=\chi_i\} Pr\{Y=\psi_j\vert X=\chi_i\},
\end{equation}

with $Pr\{Y\! =\! \psi_{j}\}$ the $j^{th}$ element of the probability mass function $p(y)$, and $Pr\{X\! =\! \chi_i\}$ the $i^{th}$ element of the pmf $p(x)$. The transmission of data over a DMC transforms the pmf of the input into the pmf of the output via a linear transformation. The probability transition matrix $Pr\{Y=\psi_j\vert X=\chi_i\}$ fully characterizes the DMC \cite{ThomasCover}. Assuming a fixed (e.g. lexicographic) order of the alphabet elements, we can introduce an index notation for the pmfs, e.g, $p^j\! :=\! Pr\{Y\! =\! \psi_{j}\}$ and $p^i\! :=\! Pr\{X\! =\! \chi_{i}\}$. In this article we associate every index with a specific random variable. In Table \ref{tbl:Indices} an overview is given.

\begin{table}[H]
\caption{Overview of Indices Used.}
\centering
\begin{tabular}{cccccc}
\toprule
 \scriptsize\textbf{Process} & \scriptsize\textbf{Variable}\normalsize &  \scriptsize\textbf{Alphabet element}\normalsize & \scriptsize\textbf{Index (input)}\normalsize & \scriptsize\textbf{Index (past)}\normalsize & \scriptsize\textbf{Index (output)}\normalsize \\
\midrule
 $X$ & $x$ & $\chi$ & $\hat{i}$  & $f$ & ${i}$\\ 
 $Y$ & $y$ & $\psi$ & $\hat{j}$ & $g$ & ${j}$\\
 $Z$ & $z$ & $\zeta$ & $\hat{k}$ & $h$ & ${k}$\\ 
\bottomrule
\end{tabular}
\label{tbl:Indices}
\end{table}

\subsection{Tensor Representation of the Communication Channel}
One of the many virtues of information theory is that it enables the use of linear algebra. Because we do not want to get overwhelmed by increasingly complex probabilistic equations, we use index notation and the Einstein summation convention (with a minor twist). Equation (\ref{eq:LawOfTotalProbability}) can now be written as $p^j = p^i A^{j}_i$. The covariant indices indicate the variables we condition on. The row stochastic probability transition matrix elements represent the elements of the probability transition tensor $\mathsf{A}$ \cite{ProbTensor}. Using the standard notation instead of the Einstein summation convention, we can rewrite MI as $I(X;Y) = \sum_{i,j} p^{ij} \log_2 \left[ A^{j}_i\big/ p^{j} \right]$.

Mutual information solely depends on the elements of the tensor and the input pmf. This is problematic in case MI or MI derived measures are used to infer the underlying structure if we assume that the structure is independent of the input. We can illustrate this by assuming that the probability transition tensor equals the Kronecker delta 

\begin{equation*} \label{eq:delta}
\delta^{j}_{i} =
    \begin{cases}
            1, &         \text{if } i=j,\\
            0, &         \text{if } i\neq j.
    \end{cases}
\end{equation*}

\begin{Example}
If $ A^{j}_i\! = \! \delta^{j}_{i}$, the symbol received is identical to the symbol sent, the channel transmits data perfectly. In this case MI reduces to $I(X;Y) = \sum_{i} p^i \log_2 \left[ \small{1} \normalsize \big/ p^{i} \right]$. Now set the probability of one of the alphabet elements to $1-\varepsilon$. This implies that all other symbol probabilities are equal to or smaller than $\varepsilon$. Taking the limit $\varepsilon \rightarrow 0$ results in a mutual information $\rightarrow 0$. Although there might be a  noiseless channel representing the association between the random variables $X$ and $Y$, MI could be arbitrarily small.
\end{Example}

This leads us to the following proposition for inferring structures using MI-based measures:

\begin{Proposition} \label{thm:structure}
 In case MI or MI related measures are used to infer the \textbf{structure} for a system, we should use the probability transition tensors or measures based on elements of probability transition tensor. 
\end{Proposition}

The earlier mentioned channel capacity is such a measure. It only depends on the elements of the probability transition tensor \cite{Muroga}, e.g. $C_{XY}\! :=\! \Gamma(\mathsf{A})$. Because the channel capacity is the maximal achievable mutual information for a specific channel, the earlier mentioned DPI also applies to the channel capacity. The proof is straightforward and therefore omitted.

\begin{Corollary}[DPI for Channel Capacity]\label{thm:DPI_ChannelCap} 
For the chain $X\! \rightarrow\! Y\! \rightarrow Z$ the DPI immediately implies that $\Gamma(\mathsf{C}) \! \leq \! \min [\Gamma(\mathsf{A}) ,\Gamma(\mathsf{B})]$, with $\mathsf{A}$ representing the tensor of the transmission $X\!\rightarrow \! Y$, $\mathsf{B}\!:\! Y\!\rightarrow \! Z$, and $\mathsf{C}\!:\! X\!\rightarrow \! Z$.
\end{Corollary}

In this short and incomplete introduction to information theory, no assumptions---other than stationarity, ergodicity and Markov property---were made about the underlying mechanisms leading to the association between random variables. We can therefore apply information theory to all cases where observational data are available.

\section{Transfer Entropy} \label{section:Transfer entropy}
Schreiber introduced transfer entropy in 2000 \cite{Schreiber}. Like MI it is non-parametric, but unlike MI it is an essentially asymmetric measure and it enables the differentiation between a source and a destination. It is an information theoretical implementation of Wieners principle of Causality \cite{Wiener}: a cause combined with the past of the effect predicts the effect better than that the effect predicts itself. In contrast to Granger causality \cite{Grangercausality}, transfer entropy can capture nonlinear relationships. We use a slightly modified version which was shown to comply fully with Wieners principle of Causality by Wibral et al. They proved that this modified TE is maximal for the real interaction delay \cite{Wibral}. We assume that $Y$ is a Markov process of order $\ell\! \geq\! 1$. This implies that the future $y_t$ also depends on its past $\textbf{y}^-\!=\! (y_{t-1}, \cdots, y_{t-\ell})$. The destination also depends on the source data $X$. With $\tau$ the finite interaction delay, we assume that for the input symbol $\textbf{x}^-\!=\! (x_{t-\tau}, \cdots, x_{t-\tau-m})$, with $m\! \geq\! 0$. The alphabet for the past of $Y$ is $\mathcal{Y}^\ell$. The alphabet for the input is $\mathcal{X}^m$.
 
\begin{equation} \label{eq:TE}
TE_{X\rightarrow Y} = \sum_{\substack{\textbf{x}^-\in \mathcal{X}^m, y\in \mathcal{Y}\\
                  \textbf{y}^-\in \mathcal{Y}^{\ell}}\\}
                  p(\textbf{x}^-,y,\textbf{y}^-) \log_2 \left[\frac{p(y|\textbf{x}^-,\textbf{y}^-)}{p(y|\textbf{y}^-)}\right]
\end{equation} 

To differentiate a source from a destination, we have to assess two hypotheses: (1) $X$ is the source and $Y$ is the destination, and (2) $Y$ is the source and $X$ is the destination. Per case the interaction delay that maximizes the respective TE is determined. If the resulting transfer entropy equals 0, we assume that there is no relation. If the TE values are larger than 0, there are in practice two possibilities: (1) the optimal interaction delays are equal: we assume that the hypothesis resulting in the largest TE is valid. (2) The optimal interaction delays are different: both hypotheses are valid so we have detected a cycle. Without loss of generality, we assume in this article that there are no cycles.

Transfer entropy is a conditional mutual information \cite{Schreiber}. Therefore, it can be associated with communication channels. We start with conditioning the MI from Equation (\ref{eq:MI}) on the event $\textbf{y}^-\!=\! \psi^-_g$ resulting in

\begin{equation*} \label{eq:MI_S}
  I(X;Y|\psi^-_g) = \sum_{\substack{\textbf{x}^-\in \mathcal{X}^m\\
                  y\in \mathcal{Y}}\\} p(\textbf{x}^-,y|\psi^-_g) \log_2 \left[ \frac{p(y|\textbf{x}^-,\psi^-_g)}{p(y|\psi^-_g)} \right].
\end{equation*}

Because $\textbf{x}^-$ and $\textbf{y}^-$ are the only parents of the output $y$, it follows from the Causal Markov Condition that the associated channel is memoryless. This mutual information quantifies the amount of information that is transmitted over the $g^{th}$ sub-channel. The transfer entropy from Equation (\ref{eq:TE}) can now be expressed as
 
\begin{equation} \label{eq:TE_Multi}
 TE_{X\rightarrow Y} = \sum_{\psi^-_g\in \mathcal{Y}^{\ell}} p(\psi^-_g) I(X;Y|\psi^-_g).
\end{equation}

We now show that this expression results from transmission of information over a set of communication channels.

\subsection{The Causal Channel}
An inverse multiplexer comprises a demultiplexer and a multiplexer in series. A demultiplexer separates an input data stream into multiple output data streams. We call these different streams sub-channels. A multiplexer combines or multiplexes all input data streams into a single output data stream \cite{MUX}.

\begin{figure}[H]
\centering
\includegraphics[width=12 cm]{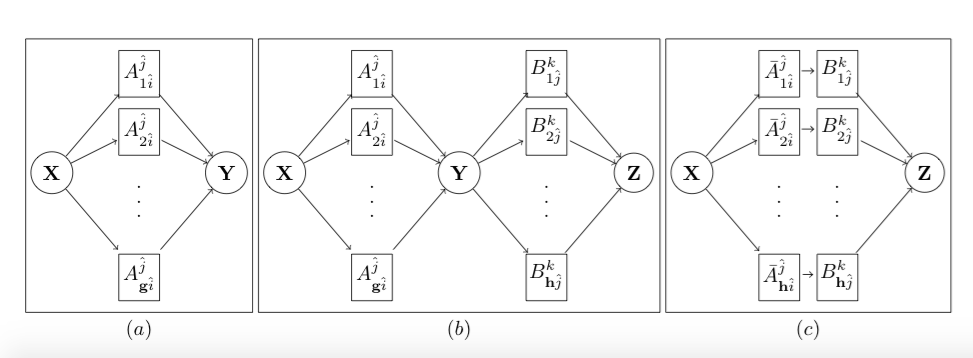}
 \caption{(\textbf{a}) Causal channel. (\textbf{b}) Two causal channels in series representing the communication model related to transfer entropy for the cascade $X\! \rightarrow\! Y\! \rightarrow\! Z$. (\textbf{c}) The equivalent causal channel for two causal channels in series.} 
\label{fig:Cascade_IM}
\end{figure}

\begin{Definition}[Causal Channel]
A causal channel is an inverse multiplexer in which the demultiplexer selects the sub-channel over which the data are send based on the past of the output data. Each sub-channel consists of a DMC. The input symbol is fed to a specific input vertex of the chosen discrete memoryless channel. The DMC transforms the input in a probabilistic fashion into an output symbol. The multiplexer combines the outputted symbols into the output message. See Figure \ref{fig:Cascade_IM}(\textbf{a}).
\end{Definition}

This definition forms the basis for the theorem that is central to this article.

\begin{Theorem}[] \label{thm:IMT}
 Transfer entropy is the average conditional mutual information of transmission over a causal channel.
\end{Theorem}

\begin{proof}
The relative frequency with which the $g^{th}$ sub-channel is chosen equals $p(\psi^-_g)$. Each sub-channel is a DMC, so the mutual information of the $g^{th}$ sub-channel equals $I(X;Y|\psi^-_g)$. The weighted average of the mutual information over all the sub-channels is equal to $\sum_{\psi^-_g\in \mathcal{Y}^{\ell}} p(\psi^-_g) I(X;Y|\psi^-_g)$, which is the definition of TE in Equation (\ref{eq:TE_Multi}).
\end{proof}

Because a DMC is a causal channel with only one sub-channel, we call a DMC a \emph{single-channel causal channel}.

\subsection{Tensor Representation of a Causal Channel}
Because every sub-channel of the causal channel represents a DMC, a causal channel can be represented by a probability transition tensor. We will call this tensor a \emph{causal tensor}. For the relation $X\rightarrow Y$ we get the following equation for the $g^{th}$ sub-channel

\begin{equation} \label{eq:XY}
 p^{j}_{g} = p^{\hat{i}}_{g} A^{j}_{g\hat{i}}.
\end{equation}

The elements of the tensor $\mathsf{A}$ are given by $A^{j}_{g\hat{i}}\! =\! p(\psi_j|\chi^{-}_{\hat{i}},\psi^{-}_{g})$. We can now rewrite TE as
 
\begin{equation} \label{eq:TE_Tensor}
 TE_{X\rightarrow Y} = \sum_{g,\hat{i},j} p^{\hat{i}jg} \log_2 \left[ \frac{A^{j}_{g\hat{i}}}{p^{j}_{g}} \right].
\end{equation}

In a similar fashion as MI, we can show that TE can be made arbitrarily close to 0 while the causal tensor itself represents a noiseless transmission. It is therefore not an optimal measure to infer structures. Again we would prefer to use the tensors themselves or measures based on these tensors like the channel capacity. The determination of the channel capacity for a causal channel is not in the scope of this article. We assume however that it is possible to determine the channel capacity.

As stated in the introduction, the approach in the article was inspired by Turing machines. The causal tensor is a realization of the transition function of a Turing machine that encodes causality in as far as the causality is encoded in the pmfs. To warrant the use of the adjective ``causal’’ however, we have to show that within the framework of causal tensors; we can differentiate between direct and indirect associations. That this seems possible can be intuited when considering the chain $X \rightarrow Y \rightarrow Z$ (see Figure \ref{fig:basic}a). The relation $X \rightarrow Z$ is a resultant of the other relations, i.e., an indirect association. Within the framework of causal tensors, we would expect that we can express this indirect association in terms of the tensors of the other relations.

\section{Calculation Rules for Causal Tensors} \label{section: calculation rules}
\begin{figure}[H] 
\centering
\includegraphics[width=10 cm]{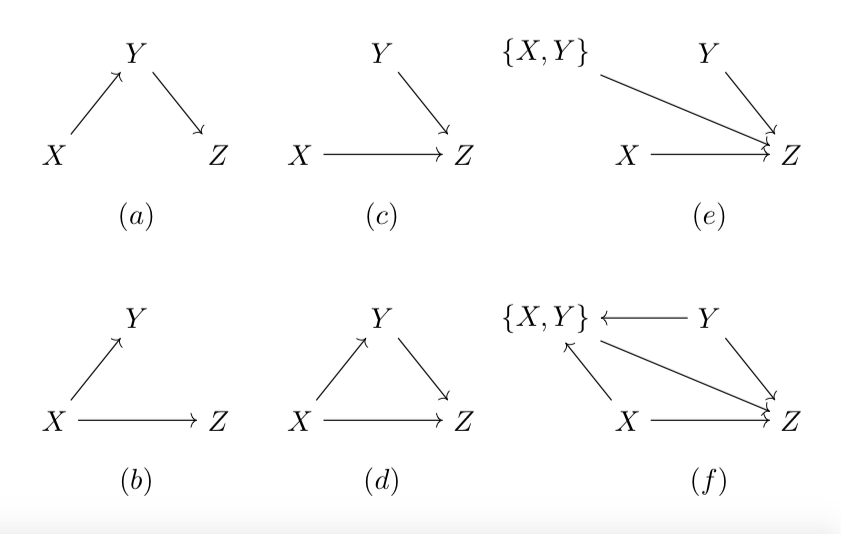} 
 \caption{\label{fig:basic} The basic structures directed graph structures: (\textbf{a}) the chain, (\textbf{b}) the fork, (\textbf{c}) the v-structure, and (\textbf{d}) the directed triangle. The graphs (\textbf{e}) and (\textbf{f}) reflect the calculation rules for the causal tensors for the v-structure and directed triangle respectively.}
\end{figure}

In this section we derive the calculation rules for causal tensors. Operations performed on these tensors should result in either scalars, stochastic vectors or stochastic tensors. To prove that we end up with stochastic tensors or vectors, we will use the earlier introduced index notation, the law of total probability and the Causal Markov Condition. We derive the calculation rules by investigating the four elementary structures depicted in Figures \ref{fig:basic}(\textbf{a}), \ref{fig:basic}(\textbf{b}), \ref{fig:basic}(\textbf{c}) and \ref{fig:basic}(\textbf{d}).

In Section \ref{section: a new notation} we will propose another notation which is simpler, but it relies on the results in this section.

\subsection{The Chain Structure}
First assume the chain $X\!\rightarrow\!Y\!\rightarrow\!Z$ is the ground truth. Additional to Equation (\ref{eq:XY}), $p^{j}_{g}\!=\! p^{\hat{i}}_{g} A^{j}_{g\hat{i}}$, there are two other causal channels represented by tensors: $\mathsf{B}\!:\! Y\!\rightarrow \! Z$, and $\mathsf{C}\!:\! X\!\rightarrow \! Z$. Because it is a straightforward exercise in which we again make use of the law of total probability, we leave it to the reader to confirm that

\begin{subequations} \label{eq:YZXZ}
\begin{align}
 p^{k}_{h} &= p^{\hat{j}}_{h} B^{k}_{h\hat{j}}, \label{eq:YZ}\\
 p^{k}_{h} &= p^{\hat{i’}}_{h} C^{k}_{h\hat{i’}}. \label{eq:XZ}
\end{align}
\end{subequations}

In principle the index $\hat{i’}$ in Equation (\ref{eq:XZ}) is the index representing a different input vector than the index $\hat{i}$ in Equation (\ref{eq:XY}); although they both refer to the random variable $X$. This is because $\hat{i’}$ is related to the source ${\textbf{x}}’^-\in \mathcal{X}^{m’}$ of $Z$ and $\hat{i}$ is the index related to the source $\textbf{x}^-\in \mathcal{X}^{m}$ of $Y$. The Markov property however immediately implies that we can use one and the same index in both cases if we select the source vector with the largest cardinality. Without loss of generality, we use $\hat{i}$.

We can express the tensor of the indirect relation in terms of the tensors of the direct relations. 

\begin{Theorem}[Product Rule for a Chain] \label{thm:CTC}
Let $\mathsf{A}$ and $\mathsf{B}$ be the causal tensors of two causal channels in series. Let the tensor $\mathsf{C}$ represent the resulting indirect causal channel that must be measured in a bivariate approach. The tensor elements of $\mathsf{C}$ are given by

\begin{equation} \label{eq:ChainTransmission}
 C^{k}_{h\hat{i}} = p^g_{h\hat{i}} A^{\hat{j}}_{g\hat{i}} B^{k}_{h\hat{j}}.
\end{equation}

If the ground truth is a directed triangle, the product rule for a chain is invalid, i.e., $C^{k}_{h\hat{i}}\! \neq\! p^g_{h\hat{i}} A^{\hat{j}}_{g\hat{i}} B^{k}_{h\hat{j}}$. 

\end{Theorem}

The proof is given in Appendix \ref{App:CTC}. The term $p^g_{h\hat{i}} A^{\hat{j}}_{g\hat{i}}$ is very interesting. In Appendix \ref{App:Averaging} we prove that it is a stochastic tensor.

\begin{Lemma} \label{eq:Averaging}
 For a chain the product $p^g_{h\hat{i}}A_{g\hat{i}}^{\hat{j}}$ is a stochastic tensor $\bar{A}^{\hat{j}}_{h\hat{i}}$.
\end{Lemma}

The causal tensor $p^g_{h\hat{i}} A^{\hat{j}}_{g\hat{i}}$ can be interpreted as the weighted average of the causal tensor $\mathsf{A}$, given the $h^{th}$ sub-channel of the final causal channel and the input $\hat{i}$. We can now rewrite Equation (\ref{eq:ChainTransmission}) as

\begin{equation} \label{eq:ChainTransmission2}
 C_{h\hat{i}}^{k} = \bar{A}^{\hat{j}}_{h\hat{i}} B_{h\hat{j}}^{k}.
\end{equation}

If both $\mathsf{A}$ and $\mathsf{B}$ represent discrete memoryless channels we get the simpler, well known, product rule for a chain of DMC’s.

\begin{Corollary}[Product Rule for a Chain of DMC’s] \label{thm:DMC}
 Let $\mathsf{A}$ and $\mathsf{B}$ be the causal tensors of two DMC’s in series and let the tensor $\mathsf{C}$ represent the resulting, indirect, causal channel that must be measured in a bivariate approach. The tensor elements of $\mathsf{C}$ are given by

\begin{equation} \label{eq:ChainTransmission_DMC}
 C^{k}_{\hat{i}} = A^{\hat{j}}_{\hat{i}} B^{k}_{\hat{j}}.
\end{equation}

\end{Corollary}

The proof follows directly from the definition of a DMC in terms of a causal channel: a discrete memoryless channel is a causal channel comprising only one sub-channel, i.e., it is a single-channel causal channel. Combined with Lemma \ref{eq:Averaging}, this corollary leads to a very specific interpretation of Equation (\ref{eq:ChainTransmission}). According to Corollary \ref{thm:DMC}, Equation \ref{eq:ChainTransmission2} can be interpreted as representing two DMC’s in series for the $h^{th}$ sub-channel. This means we have an alternative structure for two causal channels in series as depicted in Figure \ref{fig:Cascade_IM}(\textbf{c}). 

Because the Data Processing Inequality applies to a cascade of discrete memoryless channels, the alternative structure suggests that there is a DPI for transfer entropy. In Section \ref{section: structures with Causal Tensors} we show that this is indeed the case. If one so wishes we could check if the measured TE for the potential spurious association equals the expected TE. For the chain $X\rightarrow\! Y\! \rightarrow Z$ the expected $TE_{X\rightarrow Z}$ is given by

\begin{equation*} \label{eq:TE_Tensor_MICC}
 TE_{X\rightarrow Z} = \sum_{\hat{i},h,k} p^{\hat{i}hk} \log_2 \left[ \frac{\sum_{\hat{j}} \bar{A}^{\hat{j}}_{h\hat{i}} B_{h\hat{j}}^{k}}{p^{k}_{h}} \right].
\end{equation*}

\subsection{The Fork Structure}
In this section, we show that a fork can be interpreted as a chain. The product rule for a chain is therefore also applicable to a fork. Assume that the fork is the ground truth (Figure \ref{fig:basic}(\textbf{d})). Again we want to express the indirect association represented by ${\mathsf{B}}$ in terms of the other causal tensors. First, we notice that the input distribution can be reconstructed from the output distribution.

\begin{Definition}[Reconstruction Operator] \label{thm:Reconstruction}
The $\ddagger$-operator, or reconstruction operator, reconstructs the source distribution, conditioned of the past of the destination, from the destination distribution, conditioned of the past of the destination:

\begin{equation} \label{eq:XYddagger}
 p^{\hat{i}}_{g} = p^{j}_{g} A^{\ddagger \hat{i}}_{gj}, 
\end{equation}
 
with $A^{\ddagger \hat{i}}_{gj}\! =\! p^{\hat{i}}_{gj}$. The $\ddagger$-operation changes the sign of the interaction delay of the original relation. 
\end{Definition}

This implies that the directed graph $X\! \rightarrow\! Y$, is equivalent to the graph $X\! \leftarrow^{\ddagger}\! Y$. Because it is straightforward using Equation (\ref{eq:TE_Tensor}) we leave it to the reader to prove the following corollary.

\begin{Corollary}
 From an information theory point of view, a relation and its reconstructed relation are equivalent, i.e., $TE_{X \rightarrow Y} = TE_{X \leftarrow^{\ddagger} Y}$
\end{Corollary}
From this corollary immediately follows that a fork is equivalent to a chain.

\begin{Theorem}[Fork-Chain Equivalence] \label{thm:ECF}
The fork $X\! \rightarrow\! Y\! + \! X\! \rightarrow\! Z$ is equivalent to the chain $Y^{\ddagger}\! \rightarrow\! X\! \rightarrow\! Z$ and to the chain $Y\! \leftarrow\! X\! \leftarrow^{\ddagger}\! Z$.
\end{Theorem}

The indirect association represented by $\mathsf{B}$ in terms of the other two tensors of the chain follows directly from the product rule for a chain (Theorem \ref{thm:CTC}).

\begin{subequations} \label{eq:ForkTransmission}
\begin{align}
B_{h\hat{j}}^{k} &= \bar{A}^{\ddagger \hat{i}}_{h\hat{j}} C_{h\hat{i}}^{k}\text{, with } \bar{A}^{\ddagger \hat{i}}_{h\hat{j}} := p^g_{h\hat{j}}A_{g\hat{j}}^{\ddagger \hat{i}},\label{eq:ForkTransmissionBYZ}\\
B_{g\hat{k}}^{j} &= \bar{C}^{\ddagger \hat{i}}_{g\hat{k}} A_{g\hat{i}}^{j}\text{, with } \bar{C}^{\ddagger \hat{i}}_{g\hat{k}} := p^h_{g\hat{k}}{C}^{\ddagger \hat{i}}_{h\hat{k}}.\label{eq:ForkTransmissionBZY}
\end{align}
\end{subequations}

Equation (\ref{eq:ForkTransmissionBYZ}) applies in case the equivalent chain is $Y\! \rightarrow^{\ddagger}\! X\! \rightarrow\! Z$. If the equivalent chain is $Y\! \leftarrow\! X\! \leftarrow^{\ddagger}\! Z$, Equation (\ref{eq:ForkTransmissionBZY}) is applicable. Due to the way we determine the interaction delay, the $\ddagger$-operation induces a sign change for the interaction delay. E.g., if $\tau_{xy}$ represents the interaction delay for the relation $X\! \rightarrow\! Y$, then $-\tau_{xy}$ represents the interaction delay for the relation $Y^{\ddagger}\! \rightarrow\! X$. 

Matrices, and therefore tensors, do not commute: $\bar{A}^{\ddagger \hat{i}}_{h\hat{j}} \bar{A}^{\hat{j}}_{h\hat{i}}\! \neq \! \bar{A}^{\hat{j}}_{h\hat{i}}\bar{A}^{\ddagger \hat{i}}_{h\hat{j}}$. This leads to the conclusion that a chain and a fork are in principle distinguishable. The reader can verify this by combining Equation (\ref{eq:ChainTransmission2}) and Equation (\ref{eq:ForkTransmissionBYZ}). Combining this with Theorem \ref{thm:CTC} leads to the conclusion that we can  differentiate between direct and indirect relations. The conditions under which it is not possible will be derived later.

\begin{Theorem} \label{th:Chain-Fork-diff}
	Using causal tensors we can differentiate between a chain, a fork and a directed triangle. If and only if the chain is the ground truth $B_{h\hat{j}}^{k} \neq \bar{A}^{\ddagger \hat{i}}_{h\hat{j}} C_{h\hat{i}}^{k}$. If and only the fork is the ground truth $C_{h\hat{i}}^{k} \neq \bar{A}^{\hat{j}}_{h\hat{i}} B_{h\hat{j}}^{k}$. If the structure is neither a chain, nor a fork, it is a directed triangle.
\end{Theorem}

In case of a single-channel causal channels, we are allowed to ignore the index indicating the sub-channels. The equations for the chain, the fork, and consequently the equations in Theorem \ref{th:Chain-Fork-diff} are all bivariate.
 
\begin{Corollary}
If we use single-channel causal tensors, bivariate time-delayed mutual information measurements can differentiate between a fork, a chain, and a directed triangle. 
\end{Corollary}

This result contradicts the current point of view \cite{DTE, Triangle}. We illustrate this with an example. Because we use single-channel causal channels, we ignore the index $h$ in the equations.

\begin{Example}
	Let the chain $X\!\rightarrow \! Y \!\rightarrow \! Z$ be the ground truth. With $\mathsf{A}\! =\! $ \small$\begin{pmatrix} 
\frac{1}{2} & \frac{1}{2} \\
1 & 0 
\end{pmatrix}$\normalsize and $\mathsf{B}\! =\! $ \small$\begin{pmatrix} 
\frac{1}{3} & \frac{2}{3} \\
0 & 1 
\end{pmatrix}$\normalsize, the indirect association is represented by the causal tensor $C^{k}_{\hat{i}} = A^{\hat{j}}_{\hat{i}} B^{k}_{\hat{j}}\! \Rightarrow\! \mathsf{C}\! =$ \small$\begin{pmatrix} 
\frac{1}{6} & \frac{5}{6} \\
\frac{1}{3} & \frac{2}{3}  
\end{pmatrix}$\normalsize. Assume that $p(x)\! =\! (\tfrac{2}{5},\tfrac{3}{5})$. The pmf for $p(y)$ equals $p(y)\! =\! p(x)\mathsf{A}$. From this follows that $p(y)\!= \! (\tfrac{4}{5},\tfrac{1}{5})$. Using the relation $p(x)\! =\! p(y)\mathsf{A}^{\ddagger}$, the reader can verify that $\mathsf{A}^{\ddagger}\! = $ \small$\begin{pmatrix} 
\frac{1}{4} & \frac{3}{4} \\
1 & 0 
\end{pmatrix}$\normalsize. Because $\bar{A}^{\ddagger \hat{i}}_{h\hat{j}} C_{h\hat{i}}^{k}\! =\! $ \small$\begin{pmatrix} 
\frac{7}{24} & \frac{17}{24} \\
\frac{1}{6} & \frac{5}{6} 
\end{pmatrix}$\normalsize. From this follows that $B_{h\hat{j}}^{k}\! \neq \! \bar{A}^{\ddagger \hat{i}}_{h\hat{j}} C_{h\hat{i}}^{k}$. The structure is that of a chain, and not that of a fork. 
\end{Example}

\subsection{The V-Structure and the Directed Triangle}
In a bivariate measurement, we will always be able to determine the ground truth correctly in the case of the v-structure depicted in Figure \ref{fig:basic}(\textbf{c}). However, investigating structures with a collider, the v-structure and the more general directed triangle, will result in the important concept of \emph{interaction}. So, let us assume that the ground truth is the directed triangle. We now have to introduce the multivariate relation $\mathsf{D}\! :\! \{ X,Y\}\! \rightarrow \!Z$. This relation leads to the additional linear transformation
 
\begin{equation*} \label{eq:Vstructure}
p^k_{h} = p^{\hat{i} \hat{j}}_h D^{k}_{h\hat{i}\hat{j}}. 
\end{equation*}

We call the tensor $\mathsf{D}$ the \emph{interaction tensor}. The tensors $\mathsf{B}$ and $\mathsf{C}$ can be expressed in terms of the tensor $\mathsf{D}$.

\begin{Lemma}[Causal Tensor Contraction] \label{thm:TensorContract}
In the case of a directed triangle, we can express the causal tensors in terms of the interaction tensor:

 \begin{subequations} \label{eq:vstructures}
  \begin{align}
  B_{h\hat{j}}^{k} &= \bar{A}^{\ddagger \hat{i}}_{h\hat{j}} D^{k}_{h\hat{i}\hat{j}}, \label{eq:vstructure1}\\
  C_{h\hat{i}}^{k} &= \bar{A}^{\hat{j}}_{h\hat{i}} D^{k}_{h\hat{i}\hat{j}}. \label{eq:vstructure2}
  \end{align}
 \end{subequations}
   
\end{Lemma}

For the proof, we use the fact that the elements of a causal tensor are conditional probabilities. Due to the fork-chain equivalence, Appendix \ref{App:TensorContract} only contains the proof for the chain.

From Equation (\ref{eq:vstructures}) it follows that $\mathsf{B}$ and $\mathsf{C}$ are the result of a cascade involving $\mathsf{A}^{\ddagger}$ and $\mathsf{D}$ for $\mathsf{B}$, and $\mathsf{A}$ and $\mathsf{D}$ for $\mathsf{C}$. The graphs represented by Figures \ref{fig:basic}(\textbf{e}) and \ref{fig:basic}(\textbf{f}) support the tensor relations, $X\! \rightarrow\! \{ X,Y\} \! \rightarrow \! Z$ is equivalent to the cascade of the inverse multiplexers represented by $\mathsf{A}$ and $\mathsf{D}$ resulting in $\mathsf{C}$. Figures \ref{fig:basic}(\textbf{c}) and \ref{fig:basic}(\textbf{d}) however do not support the calculation rules for causal tensors.

\begin{Proposition}
 If a complex system contains v-structures, the causal graph must be represented by a directed hypergraph \cite{HyperGraph}. In a hypergraph, an edge connects any number of vertices. The interaction tensor corresponds to a so-called hyperedge.
\end{Proposition}

The interaction tensor describes the interaction of inputs at the v-structure. An indirect relation does not interact.
 
\begin{Theorem}[] \label{thm:IC}
The interaction tensor only depends on the direct causes, not on indirect causes. So, if and only if the chain is the ground truth $D^{k}_{h\hat{i}\hat{j}} = B_{h\hat{j}}^{k}$. If and only if the fork is the ground truth $D^{k}_{h\hat{i}\hat{j}} = C_{h\hat{i}}^{k}$.  
\end{Theorem}

\begin{proof}[Sketch of Proof]
 Let the ground truth be the chain. In that case, $X \bigCI Z \vert Y$ and $X$ is a non-effect of $Z$. The index $\hat{i}$ is associated with $X$, the index $\hat{j}$ is associated with $Y$ and the indices $h$ and $k$ are associated with $Z$. The Causal Markov Condition leads to $p^{{\hat{i}k}}_{h\hat{j}} = p^{{k}}_{h\hat{j}}p^{{\hat{i}}}_{h\hat{j}} \Leftrightarrow p^{{k}}_{h\hat{i}\hat{j}} = p^{{k}}_{h\hat{j}}$. 
\end{proof}
In this index ridden section, we have shown that we can express indirect relations in terms of the direct relations; the resulting tensors are stochastic tensors. We can now introduce a notation that simplifies our expressions.

\section{A New Notation} \label{section: a new notation}
The input for a causal tensor is a probability vector. This vector represents the \emph{probabilistic state} of the input. Let us assume that the input and output alphabets each consists of five alphabet elements. Because we place the operand before the operator, the probability vectors are row vectors. An input event $x$ could be represented by the probability vector $(0,0,1,0,0)$, i.e., the input equals the 3rd alphabet element. After the transmission over the causal channel, the probabilistic \emph{state} of the output element could, for example, be $(0,0.3,0,0.2,0.5)$, a probabilistic mix of the output alphabet elements. 

Apart from the probabilistic state, the states of the input and output events should also accommodate the \emph{temporal} relationship between the input and output, indicated by the interaction delay. We can unify both state characteristics by introducing complex state vectors and by adopting the \emph{braket} or Dirac notation used in quantum mechanics \cite{Dirac}. 

\subsection{Braket Notation and Tensor Operations}
The complex state vector, or the \emph{ket} is represented by the notation $\ket{x}$. From the viewpoint of index notation, the order of the operand and operator is irrelevant, so we chose to represent the probability component as a column vector, i.e., the transpose of the probability vector. In the case of our example this would be $(0,0,1,0,0)^T$. The temporal dependencies are taken care of via a complex number, e.g., $e^{i \omega t_x}$. The $i$ in $e^{i \omega t_x}$ is not an index, but the imaginary number $i$. The $\omega$ represents a normalizing constant and $t_x$ a time constant time related to the random variable, in this case, $X$. Each random variable has its own constant time associated with it. The ket in our example is represented by $\ket{x} = (0,0,1,0,0)^T e^{i \omega t_x}$.

We assume that the interaction delay is induced by the causal channel. We therefore associate it with the causal tensor. We represented it by $e^{i \omega \tau_{xy}}$, where $\tau_{xy}$ is the interaction delay between $X$ and $Y$. The \emph{complex} causal tensor is defined as $\mathcal{A}:=\mathsf{A}^Te^{i \omega \tau_{xy}}$. We can now rewrite Equation (\ref{eq:XY}) as $\ket{y} = \mathcal{A} \ket{x}$. In the case of our example, the ket for the output is given by $\ket{y} = (0,0.3,0,0.2,0.5)^T e^{i \omega (t_x+\tau_{xy})}$. To be able to simplify the product rule for a chain given by Equation (\ref{eq:ChainTransmission}), we define the cascading operator using the chain $X\!\rightarrow\!Y\!\rightarrow\!Z$.
\begin{Definition} The causal tensor cascading operator $\odot$ applied to a cascade of two causal tensors, $\mathcal{A}$ and $\mathcal{B}$, is defined as

\begin{equation*}
 \mathcal{B} \odot \mathcal{A}\ket{x} = \ket{z}.
\end{equation*}
 
\end{Definition}
Equation (\ref{eq:ChainTransmission2}) can now be written as 

\begin{equation} \label{eq:ChainTransmission2_Braket}
	\mathcal{C}=\mathcal{B} \odot \mathcal{A}. 
\end{equation}

Our definition implies that the interaction delay of a cascade of causal channels is the sum of all the interaction delays between the subsequent pairs making up the cascade. This is not an uncommon assumption \cite{Twente, Wollstadt}. The additivity also applies to ``daggered’’ relations.

\begin{Definition}
 
\begin{equation*}
\mathcal{A}^{\ddagger}:= (\mathsf{A}^{\ddagger})^T e^{-i \omega \tau_{xy}}. 
\end{equation*}

\end{Definition}

Apart from the fact that index ridden equations can be simplified, and independent of our historical attachment as the framework resulted from the intuition that a notation like this should be possible, it is a subject of future research if this novel notation provides additional new insights.

\subsection{Transfer Entropy and Mutual Information}
Both TE and MI can be expressed in terms of the complex tensors. First, we introduce a simplified notation for TE We write these measures as a function of pmfs, indicated by $(\cdot)$ and the respective tensor.

\begin{Corollary}

	\begin{equation*}
		TE_{X \rightarrow Y} = \Vert TE(\mathcal{A},\cdot) \Vert,
	\end{equation*}
with $\Vert \cdot \Vert$ the Euclidean norm. 
\end{Corollary}
The proof is straightforward when using Equation (\ref{eq:TE_Tensor}) and we therefore omit it. From now on we always assume that for measures like TE and MI we have to take the Euclidian norm. This allows us to write $TE_{X \rightarrow Y} = TE(\mathcal{A},\cdot)$. We use the subscripts, e.g., $h$, to indicate the slice representing the sub-channel in a multi-channel causal channel, e.g., $I(\mathcal{A}_h,\cdot)$ represents the MI for the $h^{th}$ sub-channel.

\section{Inferring Structures With Causal Tensors} \label{section: structures with Causal Tensors}
In this section, we discuss some non-trivial implications when using causal tensors to infer the causal structure from time-series data. First, we will show that a Data Processing Inequality for transfer entropy exists. Because we did not make any assumption about the cardinality of the alphabets used, this DPI is also valid for time-discrete, continuous data. We then prove that we can differentiate between a fork, a chain and a directed triangle as long as the data are noisy, but not ``perfectly noisy’’---we will define later this in this article. Finally we will establish a theorem that enables us to always use bivariate analysis within a system comprising three variables.

\subsection{Data Processing Inequality for TE}
The DPI for TE gives a sufficient condition to assess if a relation is a proper direct relation. It gives a necessary condition to detect potential indirect relations.

\begin{Theorem}[DPI for a Chain] \label{thm:DPI}
For the chain $X\! \rightarrow\! Y\! \rightarrow\!Z$ the following inequality holds

\begin{equation} \label{eq:DPI_TE}
 TE_{X\rightarrow Z} \! \leq \min \left[TE_{X\rightarrow Y}, \! TE_{Y\rightarrow Z} \right]. 
\end{equation}
	
Because the fork has equivalent chains, the DPI also applies to a fork.
\end{Theorem}

For readability, we have moved the proof to Appendix \ref{App:DPI}. Under the condition that the embedding of the cause and the effect vectors is sufficiently large, the DPI can identify potential indirect relations. Because we made no assumptions about the cardinality of the (finite) alphabets, the DPI is also valid for finite, very large alphabets.

\subsection{Differentiating Between Direct and Indirect Associations With Causal Tensors}
We have shown earlier that a fork, a chain, and a directed triangle are distinguishable. We now investigate in more detail under what conditions this is not possible.

\begin{Definition}[Perfect Noisy Relation]
 If and only if all causal tensor elements are equal, the relation is a perfect noisy relation. The related causal tensor is called the perfect noisy causal tensor.
\end{Definition}

The behavior of a perfect noisy causal tensor is straightforward and therefore left to the reader to confirm: (1) any input pmf is transformed into a uniform probability distribution, (2) the channel capacity = 0. The opposite of the perfect noisy causal tensor is the noiseless causal tensor.

\begin{Definition}[Noiseless Causal Tensor]
 The elements of a noiseless causal tensor satisfy $\forall_{h\hat{i}\hat{j}} A^{\hat{j}}_{h\hat{i}}\in\{0,1\} \cup \forall_h: \;\sum_{\hat{i}} {A}^{\hat{j}}_{h\hat{i}} = 1$ and $\sum_{\hat{j}} {A}^{\hat{j}}_{h\hat{i}} = 1$.
\end{Definition}

The reader can verify by using Equation (\ref{eq:TE_Tensor}) that for any input pmf, $TE=\log_2 \left[ \sum_{\hat{j}}1 \right]$. Because the channel capacity of a noiseless channel only depends on the number of alphabet elements, $C_{XY}\! =\! \min \left[\log_2 ( \vert \mathcal{X}^m \vert ),\log_2 (\vert \mathcal{Y}^\ell \vert ) \right]$ \cite{ThomasCover}, our definition is indeed a noiseless causal channel. An immediate consequence of the definition of a noiseless tensor is that the cardinality of the input pmf equals the cardinality of the output pmf. 

That we can differentiate between direct and indirect relations is related to noise. We proved the following theorem in Appendix \ref{App:ND}. 

\begin{Theorem} \label{thm:ND}
We cannot differentiate between direct and indirect relations if: (1) all relations are perfectly noiseless, or (2) the relations are (almost) perfectly noisy.
\end{Theorem}

\subsection{Bivariate Analysis With Causal tensors}
We now show that within our simple system of three variables, bivariate analysis suffices. We first need to determine the causal tensors representing multi-channel causal channels, after which we determine the causal tensors representing single-channel causal channels.

\begin{Theorem} \label{thm:Bivariate}
	If in a system comprising three variables, the ground truth is a chain, then the product rule for a chain is applicable to both the multi-channel causal channels and the single-channel causal channels:
	
\begin{equation}
 C_{h\hat{i}}^{k} = \bar{A}^{\hat{j}}_{h\hat{i}} B_{h\hat{j}}^{k} \Leftrightarrow  C^{k}_{\hat{i}} = A^{\hat{j}}_{\hat{i}} B^{k}_{\hat{j}}.
\end{equation}

The proof can be found in Section \ref{App:Bivariate}.	
\end{Theorem}

From the proof it immediately follows that this theorem is only valid when we use the same embedding and the same interaction delays for both the single-channel causal channels as for the multi-channel causal channels.

\subsection{Causal Inference Steps}
To complete the causal tensor framework as discussed so far, a summary of the (implicitly) proposed steps are given. We assume that: (1) the data are time equidistant, (2) $\ell$ and $m$ are determined correctly, and (3) the data are ergodic and stationary.

\begin{enumerate}
	\item Encode the data into a finite alphabet.
 	\item Determine the (bivariate) multi-channel causal tensors for a range of interaction delays.
 	\item Determine the optimal interaction delay.
 	\item Determine per relation the direction of causation.
 	\item Identify the potential indirect relations using the DPI and the additivity of interaction delays.
 	\item Determine the single-channel causal tensors for the potential indirect relations.
 	\item Use the product rule to determine if the indirect relations are indeed indirect.
 	\item If the network is used for simulation, determine the interaction tensors for all v-structures, i.e., determine the hypergraph.
\end{enumerate}

\section{Experiments} \label{section: Experiments}
We finalize this article with two experiments to illustrate that nonlinear behavior is indeed captured with causal tensors. 

\subsection{Ulam Map} \label{sub:Ulam map}
\begin{figure}[H]
\centering
\includegraphics[width=10 cm]{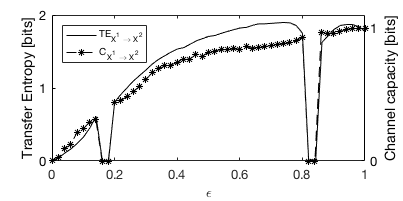} 
 \caption{\label{Ulam_Map} Transfer entropy and the channel capacity of the causal tensor for two unidirectionally coupled Ulam maps $X^{1}$ and $X^{2}$ as a function of the coupling strength $\epsilon$. Only the relation $X^{1} \rightarrow X^{2}$ is shown. Dots: approximated channel capacity for the causal channel. Line: transfer entropy as determined by Schreiber.}
\end{figure}

For the first experiment, we use the one-dimensional lattice of unidirectional coupled maps $ x^{m}_{n+1}\! =\! f \left( \epsilon x^{m-1}_{n}\! +\! (1-\epsilon) x^{m}_{n}\right)$. Information can only be transferred from $X^{m-1}$ to $X^{m}$. The Ulam map with $f(x) = 2-x^2$ is interesting because there are two regions ($\epsilon\! \approx\! 0.18$, $\epsilon\! \approx\! 0.82$) where no information is shared between maps \cite{Schreiber}. We chose an alphabet comprising four symbols. The quantization comprised simple binning. Furthermore we chose $\ell=m=1$ (see Equation (\ref{eq:TE})). Instead of maximizing TE we maximized the channel capacity to determine the optimal delay. An approximation that satisfies the boundaries that follow from Equation (\ref{eq:TE_Multi}) was used,

 \begin{equation}
  \tilde{\Gamma}(\mathsf{A}) = \sum_g p(\psi^-_g) \Gamma(\mathsf{A}_g).
 \end{equation}

To determine the channel capacities the Blahut-Arimoto algorithm was used \cite{Blahut}. The delays were varied between one and 20. The Channel capacity was maximal for a delay of one sample. As seen from Figure \ref{Ulam_Map}, causal tensors lead to a similar result as transfer entropy.

\subsection{Coupled Ornstein-Uhlenbeck Processes} \label{sub:Coupled Ornstein-Uhlenbeck processes}
\begin{figure}[H]
\centering
\includegraphics[width=8 cm]{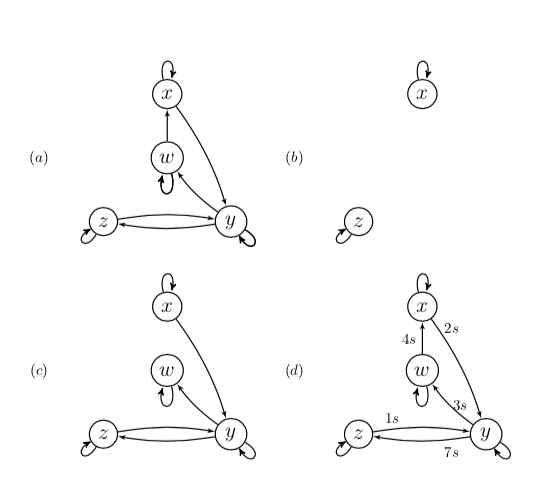} 
 \caption{\label{Potsdam} (\textbf{a}) The causal structure for the Ornstein-Uhlenbeck system of Equation (\ref{eq:Potsdam}). The other graphs show the inferred causal structures at different time series lengths. The confidence interval was $90\%$ and the maximum delay was set to $20s$: (\textbf{b}) $T=10k s$, (\textbf{c}) $T=100k s$. In (\textbf{d}), $T=500k s$, the interaction delays that maximized the channel capacity are also shown.}
\end{figure}

In the second experiment, we demonstrate our approach using a system of four coupled Ornstein-Uhlenbeck processes \cite{Runge}:
 
\begin{equation} \label{eq:Potsdam}
  \begin{aligned}
 \dot{x}(t)&=-0.5 x(t) + 0.6 w(t-4)\eta_x(t),\\  
 \dot{y}(t)&=-0.9 y(t) - 1.0 x(t-2) + 0.6z(t-5) + \eta_y(t),\\
 \dot{z}(t)&=-0.7 z(t) - 0.5 y(t-6) + \eta_z(t),\\
 \dot{w}(t)&=-0.8 w(t) - 0.4 y(t-3)^2 + 0.05 y(t-3) + \eta_w(t),
  \end{aligned}
\end{equation}

with independent unit variance white noise processes $\eta$. The integration time step was $dt=0.01s$ and the sampling interval $\Delta s=100s$. We used a binary encoding scheme. First, the data was normalized after which it was partitioned at 0.5. Because the Shannon entropy of the encoded data was close to 1, we expect highly noisy communication channels. The disadvantage of binary encoding is that more data is needed to capture the transmitted information. However, cascading very noisy channels reduces the probability of detecting an indirect relation. This is illustrated by Figure \ref{Potsdam}, no pruning was needed. This experiment shows that causal tensors can indeed detect the underlying structure.
%


\section{Discussion}
In this article we focused mainly on the basic foundations of our formalism, we assumed for example that there were no hidden causes. Whether our formalism can be used to detect the existence of hidden causes will be researched in the future. If and how our framework applies to systems comprising over three random variables will also be a subject of further research.

The conclusion that bivariate analysis suffices to differentiate direct and indirect associations, follows directly from the DPI. It should not be of any surprise that in an information theoretical framework, noise plays a fundamental role. A cascade of noisy channels is at least as noisy as the nosiest channel within the cascade.

Further experiments are needed to confirm that a combination of multi-channel causal channels and single-channel causal channels could be used for pruning inferred networks.

The experiment in Subsection \ref{sub:Ulam map} illustrated that causal tensors give similar results as transfer entropy. The experiment in Subsection \ref{sub:Coupled Ornstein-Uhlenbeck processes} illustrated that our formalism can indeed infer the underlying structure. All relations that should have been found based on the differential equations were found. All interaction delays except the interaction delay for $Z \rightarrow Y$ were close or equal to the interaction delays in the differential equations. In both cases, we used a rather simple encoding scheme. This raises the question if and under what conditions an optimal encoding scheme exists that minimizes the cardinality of the alphabets, but keeps a sufficiently large amount of ``causal information’’.

We will focus future research on the extension and application of the framework. Because in the case of a directed triangle the tensor $\mathsf{\bar{A}}$ is related to the redundancy, the \emph{potential} relation to ``Partial Information Decomposition’’ \cite{PID} will be further explored in the future.

\vspace{6pt} 



\funding{This research received no external funding.}

\acknowledgments{I would like to thank Errol Zalmijn for introducing me to the wonderful topic of transfer entropy and Marcel Brunt for helping me to implement our approach in Matlab. Also, thanks to Hans Onvlee, S. Kolumban, Rui M. Castro and T. Heskes for their comments on earlier versions of the manuscript. ASML PI System Diagnostics supported part of the work.}
\conflictsofinterest{The authors declare no conflict of interest.} 

\abbreviations{The following abbreviations are used in this manuscript:\\

\noindent 
\begin{tabular}{@{}ll}
DMC & Discrete memoryless communication channel\\
MI & Mutual information\\
pmf & probability mass function\\
TE & Transfer entropy
\end{tabular}}

\appendixtitles{yes}
\appendix
\section{Proofs}
\subsection{Theorem \ref{thm:CTC}} \label{App:CTC}
For the proof of Theorem \ref{thm:CTC} we need to introduce two lemma's.

\begin{Lemma} \label{lem:B}
	$\forall \; g:\;B^{k}_{gh\hat{j}}=B^{k}_{h\hat{j}}$.
\end{Lemma}

\begin{proof}[Sketch of Proof for Lemma \ref{lem:B}]
Another direct consequence of the Markov property is related to indices associated with the same random variable. As long as the index related to the past of the output---$g$---and the index related to the output---$j$---appear in the same tensor we are allowed to replace the output index by the input index. In our example, this means that we may replace $j$ by $\hat{j}$ as long as we ensure that $\psi^-_{\hat{j}} = \{\psi_{{j}} , \psi^-_{g}\}$. This is always possible because of the Markov property: we either enlarge the cardinality of $\psi^-_{\hat{j}}$ or $\psi^-_{g}$. 
\end{proof}
The next lemma follows directly from Example \ref{ex:CausalM}.
\begin{Lemma} \label{lem:A}
 For the chain $X\!\rightarrow\!Y\!\rightarrow\!Z$ we have $\mathcal{A}^{\hat{j}}_{\hat{i}gh}=\mathcal{A}^{\hat{j}}_{\hat{i}g}$.
\end{Lemma}

\begin{proof}[Sketch of Proof for Lemma \ref{lem:A}]
Figure \ref{Example1}(\textbf{a}) depicts the situation of the chain $X\!\rightarrow\!Y\!\rightarrow\!Z$. According to the Causal Markov Condition $\{ \hat{i}, j \}$ and $h$ are independent given $g$, i.e., $p(\hat{i}, j, h \vert g)\! =\! p(\hat{i},j\vert g)p(h\vert g)$.

We now rewrite this using expressions that follow from the definition for conditional probabilities. The left-hand side is written as $p(\hat{i}, j, h \vert g)\! =\! p(j, h \vert \hat{i},g)p(\hat{i} \vert g)$. The right-hand side can be rewritten using $p(\hat{i},j\vert g)\! =\! p(j\vert \hat{i},g)p(\hat{i}\vert g)$. 

This finally leads the conclusion that $p(j,\vert \hat{i},g,h)\! =\! p(j\vert \hat{i},g)$, i.e., $\mathcal{A}^{\hat{j}}_{\hat{i}gh}=\mathcal{A}^{\hat{j}}_{\hat{i}g}$.
\end{proof}

\begin{proof}[Sketch of Proof for Theorem \ref{thm:CTC}]
Because of the Law of Total Probability we are allowed to condition Equation (\ref{eq:XY}) on $h$ and both Equation (\ref{eq:YZ}) and Equation (\ref{eq:XZ}) on $g$. This leads to

\begin{subequations} \label{eq:YZXZnew}
\begin{align}
 p^{\hat{j}}_{gh} &= p^{\hat{i}}_{gh} A^{\hat{j}}_{gh\hat{i}},\label{eq:XYnew}\\
 p^{k}_{gh} &= p^{\hat{j}}_{gh} B^{k}_{gh\hat{j}}, \label{eq:YZnew}\\
 p^{k}_{gh} &= p^{\hat{i}}_{gh} C^{k}_{gh\hat{i}}. \label{eq:XZnew}
\end{align}
\end{subequations}

Substituting the expression for $p^{\hat{j}}_{gh}$ of Equation (\ref{eq:XYnew}) in Equation (\ref{eq:YZnew}) and combining the result with Equation (\ref{eq:XZnew}) gives us $C^{k}_{gh\hat{i}} = A^{\hat{j}}_{gh\hat{i}} B^{k}_{gh\hat{j}}$. Using Lemma(\ref{lem:B}) and Lemma(\ref{lem:A}) this can be rewritten as

\begin{equation} \label{eq:ChainTransmission_ap}
 C^{k}_{gh\hat{i}} = A^{\hat{j}}_{g\hat{i}} B^{k}_{h\hat{j}}.
\end{equation}

Finally, we multiply both sides with $p^g_{h\hat{i}}$. As the reader can confirm, the term $p^g_{h\hat{i}}C^{k}_{gh\hat{i}}$ equals $C^{k}_{h\hat{i}}$. This finally leads to Equation (\ref{eq:ChainTransmission}).\\

For the second part of the theorem, we refer to Figure \ref{Example1}(\textbf{b}). It depicts the situation of the directed triangle $X\!\rightarrow\!Y\!\rightarrow\!Z + X\!\rightarrow\!Z$. According to the Causal Markov Condition $\{ \hat{i}, j \}$ and $h$ are \emph{not} independent given $g$: $p(\hat{i}, j, h \vert g)\! \neq \! p(\hat{i},j\vert g)p(h\vert g)$. 

We now rewrite this using expressions that follow from the definition for conditional probabilities. The left-hand side is written as $p(\hat{i}, j, h \vert g)\! =\! p(j, h \vert \hat{i},g)p(\hat{i} \vert g)$. The right-hand side can be rewritten using $p(\hat{i},j\vert g)\! =\! p(j\vert \hat{i},g)p(\hat{i}\vert g)$. 
\end{proof}
\subsection{Lemma \ref{eq:Averaging}} \label{App:Averaging}

\begin{proof}[Sketch of Proof for Lemma \ref{eq:Averaging}]
 By definition $p^g_{h\hat{i}}A_{g\hat{i}}^{\hat{j}} = \sum_g p(g\vert h, \hat{i}) p(\hat{j}\vert g,\hat{i})$. From Example \ref{ex:CausalM}(\textbf{b}) it follows that $\sum_g p(g\vert h, \hat{i}) p(\hat{j}\vert g,\hat{i}) = \sum_g p(\hat{j},g\vert \hat{i},h)$. 

 Applying the law of total probability to the righthand side gives us $\sum_g p(\hat{j},g\vert \hat{i},h) = p(\hat{j}\vert \hat{i},h)$. In other words: $p^g_{h\hat{i}}A_{g\hat{i}}^{\hat{j}}=p(\hat{j}\vert \hat{i},h)$.
\end{proof}

\subsection{Lemma \ref{thm:TensorContract}} \label{App:TensorContract}
\begin{proof}[Sketch of Proof for Lemma \ref{thm:TensorContract}] \label{prf:TensorContract}
First we note that $p^{\hat{i}\hat{j}}_{h}\! =\! \delta^{\hat{j’}}_{\hat{j}} p^{\hat{j}}_h p^{\hat{i}}_{h \hat{j}’}$. Equation (\ref{eq:Vstructure}) can therefore be is rewritten as $p^k_{h} = \delta^{\hat{j’}}_{\hat{j}} p^{\hat{j}}_h p^{\hat{i}}_{h \hat{j}’} D^{k}_{h\hat{i}\hat{j}}$. Because we are allowed to change the order of $\delta^{\hat{j’}}_{\hat{j}}$ and $p^{\hat{j}}_h$ we get $p^k_{h}\! =\! p^{\hat{j}}_h \left( \delta^{\hat{j’}}_{\hat{j}} p^{\hat{i}}_{h\hat{j}’} D^{k}_{h\hat{i}\hat{j}} \right)$. Combining this with Equation (\ref{eq:YZ}) results in an expression for $B^{k}_{h\hat{j}}$: $ B^{k}_{h\hat{j}} = \delta^{\hat{j’}}_{\hat{j}} p^{\hat{i}}_{h\hat{j}’} D^{k}_{h\hat{i}\hat{j}}$. Because $\delta^{\hat{j’}}_{\hat{j}} p^{\hat{i}}_{h\hat{j}’}\! = \! p^{\hat{i}}_{h\hat{j}}$ we get Equation (\ref{eq:vstructure1}). 
\end{proof}
\subsection{Theorem \ref{thm:DPI}} \label{App:DPI}
For the proof of the data processing inequality, Theorem \ref{thm:DPI}, the simplified notation for transfer entropy and mutual information from Section \ref{section: a new notation} is used: $TE_{X \rightarrow Y}\! :=\! TE(\mathcal{A},\cdot),\; TE_{Y \rightarrow Z}\! :=\! TE(\mathcal{B},\cdot),\; TE_{X \rightarrow Z}\! :=\! TE(\mathcal{C},\cdot)$ and $I(X;Y)\! :=\! I(\mathcal{A}_h,\cdot),\; I(Y;Z)\! :=\! I(\mathcal{B}_h,\cdot),\; I(X;Z)\! :=\! I(\mathcal{C}_h,\cdot)$. The subscript $h$ indicates the $h^{th}$ sub-channel representing a DMC.

\begin{proof}[Sketch of Proof for Theorem \ref{thm:DPI}]
We start with Equation (\ref{eq:ChainTransmission2}) instead of Equation (\ref{eq:ChainTransmission2_Braket}). The DPI is valid per sub-channel. So, for all $h$: $I(\mathsf{C}_h,\cdot) \leq \min [I(\mathsf{A}_h,\cdot),I(\mathsf{B}_h,\cdot)]$. As per Equation (\ref{eq:TE_Multi}) we multiply both sides by $p(\zeta^-_h)$---the probability that the $h^{th}$ channel is selected---and sum over $h$. This results in a DPI for transfer entropy,

\begin{equation} \label{eq:thmDPI1}
 TE(\mathcal{C},\cdot) \! \leq \! \min [TE(\mathsf{\bar{A}},\cdot), TE(\mathsf{B},\cdot)].
\end{equation}

The tensor $\bar{\mathsf{A}}_h$ is itself the result of two cascaded channels represented by $\mathsf{A}_g$ and a tensor with elements $p^g_{\hat{i}h}$. For these two DMC’s the DPI is also valid, leading to: 
 
\begin{equation*}
   \forall_{g,h}:\; I(\bar{\mathsf{A}}_h,\cdot) \leq I(\mathsf{A}_g,\cdot).
\end{equation*}

We now multiply both sides of this equation by $p(\zeta^-_h)p(\psi^-_g)$, and sum over $h$ and $g$, resulting in $TE(\bar{\mathsf{A}},\cdot) \leq TE(\mathsf{A},\cdot)$. We can now rewrite Equation (\ref{eq:thmDPI1}) as

\begin{equation} \label{eq:thmDPI2}
 TE(\mathsf{C},\cdot) \! \leq \! \min [TE(\mathsf{A},\cdot), TE(\mathsf{B},\cdot)].
\end{equation}

This implies that

\begin{equation} \label{eq:thmDPI3}
 TE(\mathcal{B}\odot \mathcal{A} ,\cdot) \! \leq \! \min [TE(\mathcal{A},\cdot), TE(\mathcal{B},\cdot)].
\end{equation}
\end{proof}

\subsection{Theorem \ref{thm:ND}} \label{App:ND}

\begin{proof}[Sketch of the proof of Theorem \ref{thm:ND}]
If both $\mathcal{B} = \mathcal{C} \odot \mathcal{A}^{\ddagger}$ and $\mathcal{C} = \mathcal{B} \odot \mathcal{A}$ are valid, causal tensors can not distinguish a fork from a chain. There are two cases that need to be considered. In the first case, conditions are derived using the causal tensor relations. In the second case, we show that the pmfs impose a certain condition.

We start by combining $\mathcal{B} = \mathcal{C} \odot \mathcal{A}^{\ddagger}$ and $\mathcal{C} = \mathcal{B} \odot \mathcal{A}$:

\begin{subequations}
\label{eq:Noise}
\begin{align}
 \mathcal{B} &= \mathcal{B} \odot \mathcal{A} \odot \mathcal{A}^{\ddagger}, \label{eq:noiselessA}\\
 \mathcal{C} &= \mathcal{C} \odot \mathcal{A}^{\ddagger} \odot \mathcal{A}. \label{eq:noiselessAbar}
\end{align}
\end{subequations}

These equations are valid when $\mathcal{B} \odot \mathcal{I}_1 = \mathcal{B} \odot \mathcal{A} \odot \mathcal{A}^{\ddagger}$ and $\mathcal{C} \odot \mathcal{I}_2=\mathcal{C} \odot \mathcal{A}^{\ddagger} \odot \mathcal{A}$, with $\mathcal{I}_1$ and $\mathcal{I}_2$ identity causal tensors. Per definition identity tensors are noiseless. Because the causal tensors are stochastic tensors, their elements are nonnegative. The product of two stochastic tensors can only equal a noiseless tensor if and only if both $\mathsf{\bar{A}}$ and $\mathsf{\bar{A}}^{\ddagger}$ are noiseless. Along the same line of reasoning, we finally conclude that $\mathsf{{A}}$ and $\mathsf{{A}}^{\ddagger}$ are noiseless causal tensors because the averaging operation is in fact a matrix multiplication of two tensors.\\

The second case in which we cannot distinguish a fork and a chain follows from the pmf transformations:

\begin{subequations}
\label{eq:Noise2}
\begin{align}
 \mathcal{B}\ket{y} &= \mathcal{C}\odot \mathcal{{A}}^{\ddagger}\ket{y},\\
 \mathcal{C}\ket{x} &= \mathcal{B} \odot \mathcal{A}\ket{x}.
\end{align}
\end{subequations}

The output from both the left-hand side and the right-hand side of these equations are probability mass functions. If they are indistinguishable, we cannot differentiate between a fork and a chain either. If both $\mathcal{B}$ and $\mathcal{C}$ are (almost) perfectly noisy causal tensors, both equations in Equation (\ref{eq:Noise2}) reduce to $\ket{u} = \ket{u}$, with $\ket{u}$ representing the uniform pmf. We cannot distinguish a chain from a fork.
\end{proof}

\subsection{Theorem \ref{thm:Bivariate}} \label{App:Bivariate}
\begin{figure}[h]
\centering
\includegraphics[width=5 cm]{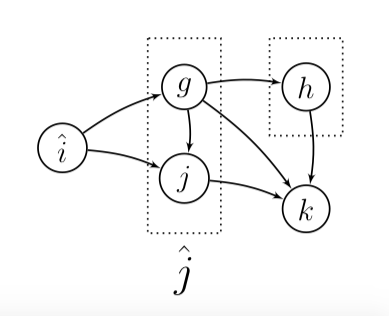}
\caption{\label{Example3} The graph related a chain of two causal channels. Both dotted boxes block the paths between $\hat{i}$ and $k$. The dotted box comprising the variables $\{g,j\}$ also blocks all paths between $\hat{i}$ and $k$. This box represents the variable $\hat{j}$.}
\end{figure}
The proof for Theorem \ref{thm:Bivariate} uses Figure \ref{Example3}. We furthermore do not use the Einstein summation convention. 

\begin{proof}[Sketch of the proof of Theorem \ref{thm:Bivariate}]
According to the Law of Total Probability, $C^{k}_{\hat{i}}\! =\! \sum_h p^h_{\hat{i}}C_{h\hat{i}}^{k}$. Multiplying both sides of Equation (\ref{eq:ChainTransmission2}) by $p^h_{\hat{i}}$, gives us 

\begin{equation} \label{a1}
 C^{k}_{\hat{i}} = \sum_{\hat{j}} \sum_h p^h_{\hat{i}} \bar{A}^{\hat{j}}_{h\hat{i}} B_{h\hat{j}}^{k}. 
\end{equation}

We now express all stochastic tensors with the letter $p$ instead of an $A$ or a $B$. Using the fact that $p^h_{\hat{i}} p^{\hat{j}}_{h\hat{i}}\! =\! p^{\hat{j}h}_{\hat{i}}$, and $p_{h\hat{j}}^{k}\! =\! p_{\hat{j}}^{kh}\big/p_{\hat{j}}^{h}$, Equation (\ref{a1}) can be written as:
 
\begin{equation} \label{a2}
 C^{k}_{\hat{i}} = \sum_{\hat{j}} \sum_h p^{\hat{j}h}_{\hat{i}} p_{\hat{j}}^{kh}\big/p_{\hat{j}}^{h}. 
\end{equation}

Because $p^{\hat{j}h}_{\hat{i}}\!=\! p^{\hat{j}}_{\hat{i}} p^{h}_{\hat{i}\hat{j}}$ and $p_{\hat{j}}^{kh}\!=\! p_{\hat{j}}^{k} p_{\hat{j}k}^{h}$ we get

\begin{equation} \label{a3}
 C^{k}_{\hat{i}} = \sum_{\hat{j}} p^{\hat{j}}_{\hat{i}} p_{\hat{j}}^{k} \sum_h p^{h}_{\hat{i}\hat{j}} p_{\hat{j}k}^{h}\big/p_{\hat{j}}^{h}. 
\end{equation}

Using $p^{h}_{\hat{i}\hat{j}}\! =\! p^{h\hat{i}}_{\hat{j}} \big/ p^{\hat{i}}_{\hat{j}}$ and $p_{\hat{j}k}^{h}\! =\! p_{\hat{j}}^{hk} \big/p_{\hat{j}}^{k}$ we rewrite Equation (\ref{a3}) as

\begin{equation} \label{a4}
 C^{k}_{\hat{i}} = \sum_{\hat{j}} p^{\hat{j}}_{\hat{i}} p_{\hat{j}}^{k} \sum_h \frac{p^{h\hat{i}}_{\hat{j}}}{p^{\hat{i}}_{\hat{j}}} \frac{p_{\hat{j}}^{hk}}{p_{\hat{j}}^{k}} \frac{1}{p_{\hat{j}}^{h}}. 
\end{equation}

Using a similar step as in the previous cases, Equation (\ref{a4}) can be rewritten as 
\begin{equation} 
 C^{k}_{\hat{i}} = \sum_{\hat{j}} p^{\hat{j}}_{\hat{i}} p_{\hat{j}}^{k} \sum_h \frac{p^{\hat{i}}_{h\hat{j}}}{p^{\hat{i}}_{\hat{j}}} \frac{p_{h\hat{j}}^{k}}{p_{\hat{j}}^{k}} p_{\hat{j}}^{h}. 
\end{equation}

Applying the Causal Markov Condition on the graph in Figure \ref{Example3}, gives us $\hat{i} \bigCI k \vert \{ h, \hat{j} \}$, i.e., $p^{\hat{i}}_{h\hat{j}}p^{k}_{h\hat{j}}p^{h}_{\hat{j}}\! =\! p^{h\hat{i}k}_{\hat{j}}$. Because we sum over $h$, we sum out this variable as per Law of Total Probability. Using $A^{\hat{j}}_{\hat{i}}=p^{\hat{j}}_{\hat{i}}$ and $B^{k}_{\hat{j}}=p^{k}_{\hat{j}}$ we finally get the following expression

\begin{equation} 
 C^{k}_{\hat{i}} = \sum_{\hat{j}} A^{\hat{j}}_{\hat{i}} B^{k}_{\hat{j}} \frac{p^{\hat{i}k}_{\hat{j}}}{p^{\hat{i}}_{\hat{j}}p^{k}_{\hat{j}}}. 
\end{equation}

So, if $\hat{i}$ and $k$ are independent given $\hat{j}$, the theorem has been proven. When we apply the Causal Markov Condition to the graph in Figure \ref{Example3}, we see that $\hat{i} \bigCI k \vert \hat{j}$.
\end{proof}
\reftitle{References}


\begin{thebibliography}{-------}
\providecommand{\natexlab}[1]{#1}

\bibitem[Guo \em{et~al.}(2018)Guo, Cheng, Li, Hahn, and Liu]{Survey}
Guo, R.; Cheng, L.; Li, J.; Hahn, P.; Liu, H.
\newblock A Survey of Learning Causality with Data: Problems and Methods {\bf
  2018}.

\bibitem[Eichler(2013)]{Eichler}
Eichler, M.
\newblock Causal inference with multiple time series: Principles and problems.
\newblock {\em Philosophical transactions. Series A, Mathematical, physical,
  and engineering sciences} {\bf 2013}, {\em 371},~20110613.
\newblock
  doi:{\changeurlcolor{black}\href{https://doi.org/10.1098/rsta.2011.0613}{\detokenize{10.1098/rsta.2011.0613}}}.

\bibitem[Pearl(2009)]{Pearl}
Pearl, J.
\newblock {\em Causality: Models, Reasoning and Inference}, 2nd ed.; Cambridge
  University Press: New York, NY, USA,  2009.

\bibitem[Granger(1969)]{GrangerORI}
Granger, C.W.J.
\newblock {Investigating Causal Relations by Econometric Models and
  Cross-Spectral Methods}.
\newblock {\em Econometrica} {\bf 1969}, {\em 37},~424--438.

\bibitem[Vastano and Swinney(1988)]{TDMI}
Vastano, J.A.; Swinney, H.L.
\newblock Information transport in spatiotemporal systems.
\newblock {\em Phys. Rev. Lett.} {\bf 1988}, {\em 60},~1773--1776.
\newblock
  doi:{\changeurlcolor{black}\href{https://doi.org/10.1103/PhysRevLett.60.1773}{\detokenize{10.1103/PhysRevLett.60.1773}}}.

\bibitem[Dagum \em{et~al.}(1992)Dagum, Galper, and Horvitz]{DBN}
Dagum, P.; Galper, A.; Horvitz, E.
\newblock Dynamic Network Models for Forecasting.
\newblock  Proceedings of the Eighth International Conference on Uncertainty in
  Artificial Intelligence; Morgan Kaufmann Publishers Inc.: San Francisco, CA,
  USA,  1992; UAI'92, pp. 41--48.

\bibitem[Spirtes \em{et~al.}(2000)Spirtes, Glymour, N., and
  Richard]{Spirtes2000}
Spirtes, P.; Glymour, C.; N., S.; Richard.
\newblock {\em Causation, Prediction, and Search}; Mit Press: Cambridge,  2000.

\bibitem[Schreiber(2000)]{Schreiber}
Schreiber, T.
\newblock Measuring Information Transfer.
\newblock {\em Phys. Rev. Lett.} {\bf 2000}, {\em 85},~461--464.
\newblock
  doi:{\changeurlcolor{black}\href{https://doi.org/10.1103/PhysRevLett.85.461}{\detokenize{10.1103/PhysRevLett.85.461}}}.

\bibitem[Lizier and Prokopenko(2010)]{Lizier2010}
Lizier, J.T.; Prokopenko, M.
\newblock Differentiating information transfer and causal effect.
\newblock {\em The European Physical Journal B} {\bf 2010}, {\em 73},~605--615.
\newblock
  doi:{\changeurlcolor{black}\href{https://doi.org/10.1140/epjb/e2010-00034-5}{\detokenize{10.1140/epjb/e2010-00034-5}}}.

\bibitem[Hyv\"{a}rinen \em{et~al.}(2010)Hyv\"{a}rinen, Zhang, Shimizu, and
  Hoyer]{ICA-LiNGAM}
Hyv\"{a}rinen, A.; Zhang, K.; Shimizu, S.; Hoyer, P.O.
\newblock Estimation of a Structural Vector Autoregression Model Using
  Non-Gaussianity.
\newblock {\em J. Mach. Learn. Res.} {\bf 2010}, {\em 11},~1709–1731.

\bibitem[Runge \em{et~al.}(2012)Runge, Heitzig, Petoukhov, and Kurths]{Runge}
Runge, J.; Heitzig, J.; Petoukhov, V.; Kurths, J.
\newblock Escaping the Curse of Dimensionality in Estimating Multivariate
  Transfer Entropy.
\newblock {\em Phys. Rev. Lett.} {\bf 2012}, {\em 108},~258701.
\newblock
  doi:{\changeurlcolor{black}\href{https://doi.org/10.1103/PhysRevLett.108.258701}{\detokenize{10.1103/PhysRevLett.108.258701}}}.

\bibitem[Duan \em{et~al.}(2013)Duan, Yang, Chen, and Shah]{DTE}
Duan, P.; Yang, F.; Chen, T.; Shah, S.
\newblock Direct Causality Detection via the Transfer Entropy Approach.
\newblock {\em Control Systems Technology, IEEE Transactions on} {\bf 2013},
  {\em 21},~2052--2066.
\newblock
  doi:{\changeurlcolor{black}\href{https://doi.org/10.1109/TCST.2012.2233476}{\detokenize{10.1109/TCST.2012.2233476}}}.

\bibitem[Sun \em{et~al.}(2014)Sun, Taylor, and Bollt]{CausationEntropy}
Sun, J.; Taylor, D.; Bollt, E.
\newblock Causal Network Inference by Optimal Causation Entropy.
\newblock {\em SIAM Journal on Applied Dynamical Systems} {\bf 2014}, {\em 14}.
\newblock
  doi:{\changeurlcolor{black}\href{https://doi.org/10.1137/140956166}{\detokenize{10.1137/140956166}}}.

\bibitem[Turing(1937)]{Turing}
Turing, A.M.
\newblock {On Computable Numbers, with an Application to the
  Entscheidungsproblem}.
\newblock {\em Proceedings of the London Mathematical Society} {\bf 1937}, {\em
  s2-42},~230--265,
  \href{http://xxx.lanl.gov/abs/http://oup.prod.sis.lan/plms/article-pdf/s2-42/1/230/4317544/s2-42-1-230.pdf}{{\normalfont
  [http://oup.prod.sis.lan/plms/article-pdf/s2-42/1/230/4317544/s2-42-1-230.pdf]}}.
\newblock
  doi:{\changeurlcolor{black}\href{https://doi.org/10.1112/plms/s2-42.1.230}{\detokenize{10.1112/plms/s2-42.1.230}}}.

\bibitem[Copeland(2019)]{Church_Turing}
Copeland, B.J.
\newblock The Church-Turing Thesis. In {\em The Stanford Encyclopedia of
  Philosophy}, Spring 2019 ed.;  Zalta, E.N., Ed.; Metaphysics Research Lab,
  Stanford University,  2019.

\bibitem[Pearl(2009)]{SCM}
Pearl, J.
\newblock Causal Inference in Statistics: An Overview.
\newblock {\em Statistics Surveys} {\bf 2009}, {\em 3},~96--146.
\newblock
  doi:{\changeurlcolor{black}\href{https://doi.org/10.1214/09-SS057}{\detokenize{10.1214/09-SS057}}}.

\bibitem[Eberhardt and Scheines(2007)]{Interventions}
Eberhardt, F.; Scheines, R.
\newblock Interventions and Causal Inference.
\newblock {\em Philos. Sci.} {\bf 2007}, {\em 74}.
\newblock
  doi:{\changeurlcolor{black}\href{https://doi.org/10.1086/525638}{\detokenize{10.1086/525638}}}.

\bibitem[Ding \em{et~al.}(2006)Ding, Chen, and Bressler]{Grangercausality}
Ding, M.; Chen, Y.; Bressler, S.L.
\newblock Granger Causality: Basic Theory and Application to Neuroscience,
  2006.

\bibitem[Ghassami and Kiyavash(2017)]{Triangle}
Ghassami, A.; Kiyavash, N.
\newblock Interaction information for causal inference: The case of directed
  triangle.
\newblock  2017 IEEE International Symposium on Information Theory (ISIT),
  2017, pp. 1326--1330.
\newblock
  doi:{\changeurlcolor{black}\href{https://doi.org/10.1109/ISIT.2017.8006744}{\detokenize{10.1109/ISIT.2017.8006744}}}.

\bibitem[Cover and Thomas(1991)]{ThomasCover}
Cover, T.M.; Thomas, J.A.
\newblock {\em {Elements of Information Theory}}; Wiley-Interscience: New York,
  NY, USA,  1991.

\bibitem[Margolin \em{et~al.}(2006)Margolin, Nemenman, Basso, Wiggins,
  Stolovitzky, Dalla-Favera, and Califano]{ARACNE_DPI}
Margolin, A.; Nemenman, I.; Basso, K.; Wiggins, C.; Stolovitzky, G.;
  Dalla-Favera, R.; Califano, A.
\newblock ARACNE: An Algorithm for the Reconstruction of Gene Regulatory
  Networks in a Mammalian Cellular Context.
\newblock {\em BMC bioinformatics} {\bf 2006}, {\em 7 Suppl 1},~S7.
\newblock
  doi:{\changeurlcolor{black}\href{https://doi.org/10.1186/1471-2105-7-S1-S7}{\detokenize{10.1186/1471-2105-7-S1-S7}}}.

\bibitem[Papoulis and Pillai(2002)]{LoTP}
Papoulis, A.; Pillai, S.U.
\newblock {\em Probability, Random Variables, and Stochastic Processes}, fourth
  ed.; McGraw Hill,  2002.

\bibitem[Shannon()]{Shannon}
Shannon, C.E.
\newblock A Mathematical Theory of Communication.
\newblock {\em Bell System Technical Journal}, {\em 27},~379--423,
  \href{http://xxx.lanl.gov/abs/https://onlinelibrary.wiley.com/doi/pdf/10.1002/j.1538-7305.1948.tb01338.x}{{\normalfont
  [https://onlinelibrary.wiley.com/doi/pdf/10.1002/j.1538-7305.1948.tb01338.x]}}.
\newblock
  doi:{\changeurlcolor{black}\href{https://doi.org/10.1002/j.1538-7305.1948.tb01338.x}{\detokenize{10.1002/j.1538-7305.1948.tb01338.x}}}.

\bibitem[{Ahmed} \em{et~al.}(2018){Ahmed}, {Roy}, and
  {Kalita}]{EffectivenessCI}
{Ahmed}, S.S.; {Roy}, S.; {Kalita}, J.K.
\newblock Assessing the Effectiveness of Causality Inference Methods for Gene
  Regulatory Networks.
\newblock {\em IEEE/ACM Transactions on Computational Biology and
  Bioinformatics} {\bf 2018}, pp. 1--1.
\newblock
  doi:{\changeurlcolor{black}\href{https://doi.org/10.1109/TCBB.2018.2853728}{\detokenize{10.1109/TCBB.2018.2853728}}}.

\bibitem[Rashidi \em{et~al.}(2018)Rashidi, Singh, and Zhao]{RASHIDI}
Rashidi, B.; Singh, D.S.; Zhao, Q.
\newblock Data-driven root-cause fault diagnosis for multivariate non-linear
  processes.
\newblock {\em Control Engineering Practice} {\bf 2018}, {\em 70},~134 -- 147.
\newblock
  doi:{\changeurlcolor{black}\href{https://doi.org/https://doi.org/10.1016/j.conengprac.2017.09.021}{\detokenize{https://doi.org/10.1016/j.conengprac.2017.09.021}}}.

\bibitem[Dullemond and Peeters(2010)]{ProbTensor}
Dullemond, K.; Peeters, K.
\newblock Introduction to Tensor Calculus.
\newblock  2010.

\bibitem[{Muroga}(1953)]{Muroga}
{Muroga}, S.
\newblock {On the Capacity of a Discrete Channel.}
\newblock {\em Journal of the Physical Society of Japan} {\bf 1953}, {\em
  8},~484--494.
\newblock
  doi:{\changeurlcolor{black}\href{https://doi.org/10.1143/JPSJ.8.484}{\detokenize{10.1143/JPSJ.8.484}}}.

\bibitem[Beckenbach(1961)]{Wiener}
Beckenbach, E.F.
\newblock {\em Modern mathematics for the engineer: second series}; New York :
  McGraw-Hill,  1961.

\bibitem[Wibral \em{et~al.}(2013)Wibral, Pampu, Priesemann, Siebenh\"uhner,
  Seiwert, Lindner, Lizier, and Vicente]{Wibral}
Wibral, M.; Pampu, N.; Priesemann, V.; Siebenh\"uhner, F.; Seiwert, H.;
  Lindner, M.; Lizier, J.T.; Vicente, R.
\newblock Measuring information-transfer delays.
\newblock {\em PloS one} {\bf 2013}.

\bibitem[Dean(2012)]{MUX}
Dean, T.
\newblock {\em Network+ Guide to Networks}, 6th ed.; Course Technology Press:
  Boston, MA, United States,  2012.

\bibitem[Berge(1985)]{HyperGraph}
Berge, C.
\newblock {\em Graphs and Hypergraphs}; Elsevier Science Ltd.: Oxford, UK, UK,
  1985.

\bibitem[Dirac(1939)]{Dirac}
Dirac, P.A.M.
\newblock A new notation for quantum mechanics.
\newblock {\em Mathematical Proceedings of the Cambridge Philosophical Society}
  {\bf 1939}, {\em 35},~416–418.
\newblock
  doi:{\changeurlcolor{black}\href{https://doi.org/10.1017/S0305004100021162}{\detokenize{10.1017/S0305004100021162}}}.

\bibitem[Nauta \em{et~al.}(2019)Nauta, Bucur, and Seifert]{Twente}
Nauta, M.; Bucur, D.; Seifert, C.
\newblock Causal Discovery with Attention-Based Convolutional Neural Networks.
\newblock {\em Machine Learning and Knowledge Extraction} {\bf 2019}, {\em
  1},~312--340.
\newblock
  doi:{\changeurlcolor{black}\href{https://doi.org/10.3390/make1010019}{\detokenize{10.3390/make1010019}}}.

\bibitem[{Wibral} \em{et~al.}(2012){Wibral}, {Wollstadt}, {Meyer}, {Pampu},
  {Priesemann}, and {Vicente}]{Wollstadt}
{Wibral}, M.; {Wollstadt}, P.; {Meyer}, U.; {Pampu}, N.; {Priesemann}, V.;
  {Vicente}, R.
\newblock Revisiting Wiener's principle of causality — interaction-delay
  reconstruction using transfer entropy and multivariate analysis on
  delay-weighted graphs.
\newblock  2012 Annual International Conference of the IEEE Engineering in
  Medicine and Biology Society,  2012, pp. 3676--3679.
\newblock
  doi:{\changeurlcolor{black}\href{https://doi.org/10.1109/EMBC.2012.6346764}{\detokenize{10.1109/EMBC.2012.6346764}}}.

\bibitem[Blahut(1972)]{Blahut}
Blahut, R.
\newblock Computation of channel capacity and rate-distortion functions.
\newblock {\em IEEE Transactions on Information Theory} {\bf 1972}, {\em
  18},~460--473.
\newblock
  doi:{\changeurlcolor{black}\href{https://doi.org/10.1109/TIT.1972.1054855}{\detokenize{10.1109/TIT.1972.1054855}}}.

\bibitem[Williams and Beer(2010)]{PID}
Williams, P.; Beer, R.
\newblock Nonnegative Decomposition of Multivariate Information.
\newblock {\em preprint} {\bf 2010}, {\em 1004}.

\end{thebibliography}
\end{document}